\documentclass[10pt,journal,compsoc]{IEEEtran}
\usepackage[utf8]{inputenc}
\usepackage{multirow}
\usepackage{array}
\usepackage{bbding}
\usepackage{graphicx}  
\usepackage{tablefootnote}
\usepackage{flushend}
\usepackage{amsmath,amsfonts}
\usepackage[colorlinks,linkcolor=black]{hyperref}

\ifCLASSOPTIONcompsoc
  \usepackage[nocompress]{cite}
\else
  \usepackage{cite}
\fi

\ifCLASSINFOpdf
\else
\fi

\begin{document}
%
\title{The Partially Observable Asynchronous Multi-Agent Cooperation Challenge}
%
%
%
%

\author{Meng Yao,
        Qiyue Yin*,
        Jun~Yang,
        Tongtong Yu,
        Shengqi Shen,
        Junge Zhang,
        Bin Liang,
        Kaiqi Huang 
\IEEEcompsocitemizethanks{\IEEEcompsocthanksitem Meng Yao, Qiyue Yin, Tongtong Yu, Shengqi Shen, Junge Zhang and Kaiqi Huang are
with Institute of Automation, Chinese Academy of Sciences, Beijing,
China, 100190; Qiyue Yin and Kaiqi Huang are also with University of Chinese Academy of Sciences,
Beijing, China, 100049.\protect\\
E-mail: meng.yao@ia.ac.cn, qyyin@nlpr.ia.ac.cn, tongtong.yu, shengqi.shen, jgzhang@nlpr.ia.ac.cn, kqhuang@nlpr.ia.ac.cn
\IEEEcompsocthanksitem Jun Yang and Bin Liang are with the Department of Automation, Tsinghua University, Beijing,
China, 100084.\protect\\
E-mail: yangjun603@tsinghua.edu.cn, bliang@tsinghua.edu.cn
\IEEEcompsocthanksitem * Corresponding author: Qiyue Yin
}
}

%
%

\markboth{Journal of \LaTeX\ Class Files,~Vol.~14, No.~8, August~2015}%
{Shell \MakeLowercase{\textit{et al.}}: Bare Demo of IEEEtran.cls for Computer Society Journals}
%



\IEEEtitleabstractindextext{%
\begin{abstract}
Multi-agent reinforcement learning (MARL) has received increasing attention for its applications in various domains. Researchers have paid much attention on its partially observable and cooperative settings for meeting real-world requirements. For testing performance of different algorithms, standardized environments are designed such as the StarCraft Multi-Agent Challenge, which is one of the most successful MARL benchmarks. To our best knowledge, most of current environments are synchronous, where agents execute actions in the same pace. However, heterogeneous agents usually have their own action spaces and there is no guarantee for actions from different agents to have the same executed cycle, which leads to asynchronous multi-agent cooperation. Inspired from the Wargame, a confrontation game between two armies abstracted from real world environment, we propose the first Partially Observable Asynchronous multi-agent Cooperation challenge (POAC) for the MARL community. Specifically, POAC supports two teams of heterogeneous agents to fight with each other, where an agent selects actions based on its own observations and cooperates asynchronously with its allies. Moreover, POAC is a light weight, flexible and easy to use environment, which can be configured by users to meet different experimental requirements such as self-play model, human-AI model and so on. Along with our benchmark, we offer six game scenarios of varying difficulties with the built-in rule-based AI as opponents. Finally, since most MARL algorithms are designed for synchronous agents, we revise several representatives to meet the asynchronous setting, and the relatively poor experimental results validate the challenge of POAC. Source code is released in \url{http://turingai.ia.ac.cn/data\_center/show}.

\end{abstract}

}

\maketitle

\IEEEdisplaynontitleabstractindextext

%
\IEEEpeerreviewmaketitle

\IEEEraisesectionheading{\section{Introduction}\label{sec:introduction}}
\IEEEPARstart{M}{any} real-world problems can be modeled as cooperation of multiple agents, such as self-driving cars \cite{selfdriving}, multi-robot control \cite{multirobot}, the networking of communication packages \cite{netcom}, and the trading on financial markets \cite{fain}.
Recently, A significant part of the research on multi-agent cooperation focuses on reinforcement learning techniques, leading to repaid development of multi-agent reinforcement learning (MARL).
Plenty of works have been developed such as \cite{berner2019dota} \cite{baker2019emergent} \cite{bansal2017emergent} \cite{jaderberg2019human} and promising performances have been obtained.
One of key factors that promotes MARL is game environments, where new MARL algorithms can be quickly tested in a safe and reproducible manner.
As one of the most successful environments, the StarCraft Multi-Agent Challenge (SMAC) \cite{smac}, concentrates on providing benchmarks for partially observable, cooperative, multi-agent learning problem, based on which, a lot of influential works have been verified.
Unlike SMAC, whose cooperation are synchronous, we propose, to the best of our knowledge, the first partially observable asynchronous multi-agent cooperation challenge (POAC) for the MARL community, which we argue is more general than synchronous cooperation in real world.

\begin{figure}
   \begin{center}
   \includegraphics[width=0.4\textwidth]{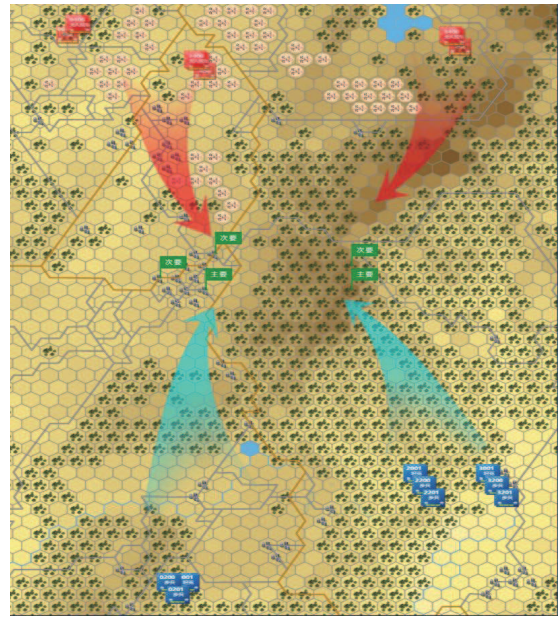}
   \end{center}
   \caption{A game instance of wargame.}
   \label{bq}
\end{figure}

POAC is inspired from Wargame, a confrontation game between two armies, where two teams of heterogeneous agents fight against each other in a partially observable environment, and most importantly cooperate asynchronously for beating enemies.
Generally, Wargame is a very complex game with plenty of rules such as moving rules, shooting rules and ruling rules.
A screenshot is shown in Figure \ref{bq}\footnote{Come from http://turingai.ia.ac.cn/}.
Directly bringing in a complete Wargame as a benchmark for testing MARL algorithms is impractical, because much effort will be wasted to learn skills unrelated with asynchronous cooperation such as exploring key points in a large map, learning to shoot with proper weapons under random adjudication.
So, we abstract POAC from Wargame by keeping the main features and dropping some specific domain knowledge to make it a universal game environment.
Specifically, POAC is an asynchronous, heterogeneous, real-time, imperfect information game with stochasticity (state transition is stochastic).

We tried our best to make POAC a light weight, flexible and easy to use benchmark:
\begin{itemize}
\item POAC can be easily configured by users to meet different experimental requirements such as switching self-play, human control and learning against built-in bot modes.
\item POAC offers six game scenarios with various built-in rule-based AIs as opponents, based on which, researchers can train their asynchronous operation strategies.
\item We revise several representative MARL algorithms to meet the asynchronous setting, and the experimental result and implementation code provide baselines for community.
\end{itemize}

\section{Related Work}
There are several multi-agent game benchmarks that have promoted the development of MARL algorithms.
Multiple gridworld-like environments \cite{lowe} are a set of simple grid-world like environments for multi-agent RL with an implementation of MADDPG for mix of competitive and cooperative tasks, which focus on shared communication and low level continuous control.
\cite{leibo} is a mixed-cooperative Markov environment aiming to test social dilemmas.
\cite{yang2018mean} focuses on testing emergent behaviour, and presents a framework for creating gridworlds covering hundreds to the millions of agents with relatively simple game rules.
\cite{resnick2018pommerman} proposes a multi-agent environment based on the game Bomberman, consisting of a series of cooperative and adversarial tasks to be more challenging but with simple grid-world-like action spaces.
The SMAC \cite{smac}, which is a representative of the challenging multi-agent game, has been used as a test-bed for various MARL algorithms and so does the recently proposed Google Research Football \cite{kurach2019google} environment.
However, we argue some of them do not provide qualitatively challenging benchmark environments for testing MARL algorithms and almost all of them ignore asynchronous cooperation, which is common tasks in real world.
Although the most recently developed Fever Basketball\cite{jia2020fever} benchmark is an asynchronous cooperation sports game environment for MARL community, such environment shares perfect information for every agents, which is usually unpractical.

\section{POCA}
\subsection{POCA environment}
\begin{figure}[!h]
   \begin{center}
   \includegraphics[width=0.4\textwidth]{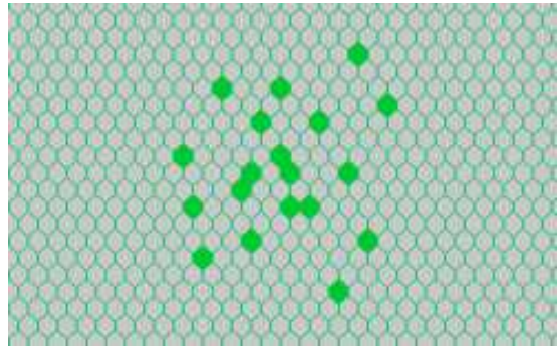}
   \end{center}
   \caption{An example of map, the solid girds are special grids with hidden terrain property.}
   \label{map}
\end{figure}

Generally, operators from red and blue teams fight against each other on a specific map by obeying pre-given fighting rules in a wargame.
Similarly, POCA consists of three basic elements: operator, map and rule.
We utilize three different types of operators: chariot, tank and infantry, with each of them having unique attributes.
We will elaborate this in the following parts.
Like Wargame, POCA uses hexagonal map for more freedom of movement, as shown in Figure \ref{map}.

\begin{figure}[!h]
   \begin{center}
   \includegraphics[width=0.4\textwidth]{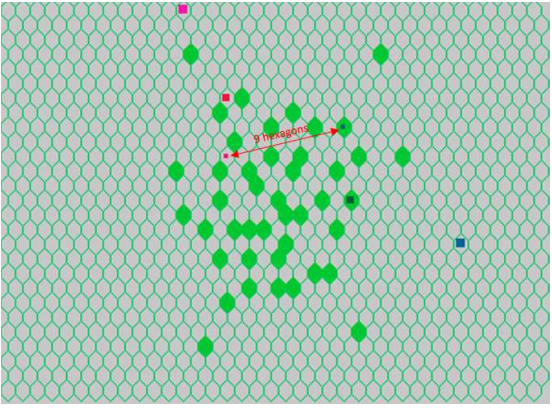}
   \end{center}
   \caption{The solid girds are special grids with hidden terrain property. For example, tank's observed distance is 10, but when tank is in a special terrain, its observed distance would be 5. The red tank (little red square) is 9 hexagons away from the blue tank (little green square), so the blue tank can see the red tank, but the red tank can't see the blue tank, because the blue tank is in a special terrain.}
   \label{featureSee}
\end{figure}

\begin{figure}[!h]
   \begin{center}
   \includegraphics[width=0.4\textwidth]{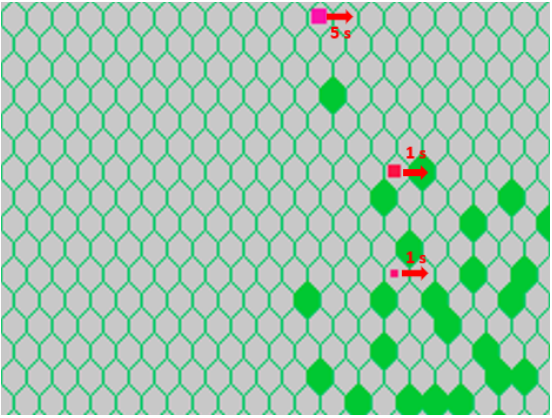}
   \end{center}
   \caption{Different moving speeds resulting in asynchronous cooperation.}
   \label{yibu}
\end{figure}

As for the rules, there are several distinct rules making POCA different from previous benchmarks.
\begin{itemize}
\item Some hexagons of the map have hidden terrain property, where it is difficult for enemies to observe when operators located in because of shortened observed distance. As shown in Figure \ref{featureSee}. Based on such a rule, partially observable feature can be reflected even when different operators are fighting in a nearby distance.
\item Move, as a very important action in POAC, requires different time steps moving to an adjacent hexagon for different operators, which is a key factor making POCA an asynchronous game. This is shown in Figure \ref{yibu}.
\item When an agent attacks another agent, it causes damage based on random number seed, making POCA a random environment, i.e., uncertain in state transition even when the state is fully observable.
\item A special cooperation rule, called guide shoot, needs two operators cooperate in a particular way so as to cause damage that may not be completed alone, which brings new challenges to MARL agent for achieving efficient cooperation.
\end{itemize}


\subsection{POCA for MARL}
POAC is a partially observable asynchronous multi-agent cooperation, designed for cooperative multi-agent tasks described as Dec-POMDPs \cite{oliehoek2016concise}.
Formally, a Dec-POMDP $G$ is given by a tuple $G=<S, U, P, r, Z, O, n, \gamma>$, and $s \in \mathrm{S}$ is the true state of the environment.
At each time step, each agent $a \in A \equiv\{1, \ldots, n\}$ chooses an action $u^{a} \in U$ and forms a joint action $\mathbf{u} \in \mathbf{U} \equiv U^{n}$, which causes a transition of the environment according to the state transition function $P\left(s^{\prime} \mid s, \mathbf{u}\right): S \times \mathbf{U} \times S \rightarrow[0,1]$.
In a partially-observable stochastic game, usually all agents in a Dec-POMDP share the same team reward $r(s, \mathbf{u}): S \times \mathbf{U} \rightarrow \mathbb{R}$, and $\gamma \in[0,1)$ is the discount factor. Dec-POMDPs consider partially observable scenarios where an observation function $O(s, a): S \times A \rightarrow Z$ determines the observations $z^{a} \in Z$ of the agent.
Each agent has an action-observation history $\tau^{a} \in T \equiv(Z \times U)^{*}$ on which it conditions a policy $\pi^{a}\left(u^{a} \mid \tau^{a}\right): T \times U \rightarrow[0,1]$.
The joint policy $\pi$ forms a joint action-value function: $Q^{\pi}\left(s_{t}, \mathbf{u}_{t}\right)=\mathbb{E}_{s_{t+1: \infty}, \mathbf{u}_{t+1: \infty}}\left[R_{t} \mid s_{t}, \mathbf{u}_{t}\right]$, and $R_{t}=\sum_{i=0}^{\infty} \gamma^{i} r_{t+i}$ is the discounted return.

In most multi-agent benchmarks, synchrony between actions is an implicit assumption.
It is called synchronous if there is a global clock and agents move in lockstep and a step in the system corresponds to a tick of the clock.
However, in an asynchronous cooperation system, there is no global clock, and the agents in the system can run at arbitrary rates relative to each other \cite{halpern2007computer}.
In POCA, heterogeneous agents have their own action spaces and executed cycle, which are different from each other.
Because of this, Dec-POMDPs should be modified for asynchronous tasks.
Specifically, at each time step (suppose a global clock is existed), each agent chooses an action $u^{a} \in A$ forming a joint action $\mathbf{u}^{val} \in \mathbf{U}^{val} $, where $\mathbf{u}^{val}$ is actions of agents that can perform actions. This will cause a transition of the environment $P\left(s^{\prime} \mid s, \mathbf{u}^{val}\right): S \times \mathbf{U}^{val} \times S \rightarrow[0,1]$.

Next we describe key factors for designing MARL algorithms on POCA, based on which, we implement representative centralised training with decentralised execution MARL algorithms, i.e., QMIX\cite{rashid2018qmix},QTRAN\cite{son2019qtran},COMA\cite{foerster2018counterfactual},VDN\cite{sunehag2017value} and IQL\cite{tampuu2017multiagent}, as baselines to test our environment.

\textbf{States and Observations}.
At each timestep, agents receive local observations within their field of view. The sight range makes the environment partially observable from the standpoint of each agent.
Agents can only observe other agents if they are both alive and located within the sight range.
In POCA, We use the attributes information of the operator as state and observation, and a general summarization is shown in Table \ref{att}.
Differen operators share very distinct characteristics, as shown in Tables \ref{tab:tank}, \ref{tab:chariot} and \ref{tab:infantry}.
From the tables, we can see that tank has the most health, no shoot preparation time, higher probability of attack damage, so tank is suitable for charging.
For chariot, it can make the most damage to tank and it is also main player for using guide shoot, but because of its shoot preparation time, it can not shoot while move, so chariot should be well protected.
At last, infantry has slowest movement speed, but it can take the most damage, and because its observed distance is so short that it can not be easily detected by the enemy, so it can hide for finding enemies and cooperate with chariot to fight with enemies using guide shoot.
\begin{table*}[htbp]
\begin{center}
\caption{Operators attributes.}
\label{att}
\begin{tabular}{l l}
\hline
Attributes     &   Description   \\
\hline
color & 0 (red) or 1 (blue)  \\
id & decimal number  \\
type & 0 for tank, 1for chariot, and 2 for infantry  \\
blood & current blood  \\
position & current position  \\
speed & move speed  \\
observed distance & max distance for an operator to be observed  \\
attacked distance & max distance for an operator to be attacked  \\
attack damage against tank or chariot  & damage obtained when attacking tank or chariot operators  \\
probability of causing damage against tank or chariot  &  probability of causing damage when attacking tank or chariot operators  \\
attack damage against infantry  & damage obtained when attacking infantry  \\
probability of causing damage against infantry  & probability of causing damage when attacking  infantry \\
shoot cool-down & cool-down time after an attack  \\
shoot preparation time & time required to attack enemy operators   \\
move time & time from the beginning of move to the present \\
stop time & time from the beginning of stop to the present \\
can guide shoot & true or false \\
\hline
\end{tabular}
\end{center}
\end{table*}
\begin{table}[htbp]
\begin{center}
\caption{Operators details for tank.}
\label{tab:tank}
\begin{tabular}{l l}
\hline
Tank     &      \\
\hline
blood & 10\\
speed & 1\\
observed distance & 10\\
attacked distance & 7\\
attack damage against tank or chariot  & 1.2  \\
probability of causing damage against tank or chariot  &  0.8  \\
attack damage against infantry  & 0.6  \\
Probability of causing damage against infantry  & 0.6 \\
shoot cool-down & 1 \\
shoot preparation time & 0   \\
can guide shoot & False \\
\hline
\end{tabular}
\end{center}
\end{table}
\begin{table}[htbp]
\begin{center}
\caption{Operators details for chariot.}
\label{tab:chariot}
\begin{tabular}{l l}
\hline
Chariot     &      \\
\hline
blood & 8\\
speed & 1\\
observed distance & 10\\
attacked distance & 7\\
attack damage against tank or chariot  & 1.5  \\
probability of causing damage against tank or chariot  &  0.7  \\
attack damage against infantry  & 0.8  \\
Probability of causing damage against infantry  & 0.6 \\
shoot cool-down & 1 \\
shoot preparation time & 2   \\
can guide shoot & True \\
\hline
\end{tabular}
\end{center}
\end{table}
\begin{table}[htbp]
\begin{center}
\caption{Operators details for infantry.}
\label{tab:infantry}
\begin{tabular}{l l}
\hline
Infantry     &      \\
\hline
blood & 7\\
speed & 0.2\\
observed distance & 5\\
attacked distance & 3\\
attack damage against tank or chariot  & 0.8  \\
probability of causing damage against tank or chariot  &  0.7  \\
attack damage against infantry  & 0.8  \\
Probability of causing damage against infantry  & 0.6 \\
shoot cool-down & 1 \\
shoot preparation time & 2   \\
can guide shoot & True \\
\hline
\end{tabular}
\end{center}
\end{table}

The feature vector observed by each agent contains important attributes for both allied and enemy units within the sight range such as faction (color), id, type, current position, blood, as summarized in Table \ref{tab:feature}.
Besides, the time information and the can\_see and can\_attack are considered.
can\_see shows enemy index within agent's sight, and can\_attack shows enemy index which can be attacked by agent.
Finally, the global state, which is only available to agents during centralised training, contains perfect information about all operators on the map.
All features, both in the state as well as in the observations of individual agents, are normalised by their maximum values.
It is worth mentioning that we provide map feature for more comprehensive information with open interface.

\begin{table}[htbp]
\begin{center}
\caption{Feature vectors for each agent.}
\label{tab:feature}
\begin{tabular}{l l l l}
allied\_feature & type & placeholder & local/global state \\
\hline
color & float & 1 & True/True\\
id & float & 1 & True/True\\
type & float & 1 & True/True\\
cur\_hex & float & 1 & True/True\\
blood & float & 1 & True/True\\
move\_time & float & 1 & True/True\\
stop\_time & float & 1 & True/True\\
shoot\_cooling\_time & float & 1 & True/True\\
can\_see & float & 3 & True/True\\
can\_attack & float & 3 & True/True\\
\hline
\end{tabular}
\begin{tabular}{l l l l}
enemy\_feature & type & placeholder & local/global state \\
\hline
color & float & 1 & True/True\\
id & float & 1 & True/True\\
type & float & 1 & True/True\\
cur\_hex & float & 1 & True/True\\
blood & float & 1 & True/True\\
move\_time & float & 1 & False/True\\
stop\_time & float & 1 & False/True\\
shoot\_cooling\_time & float & 1 & False/True\\
can\_see & float & 3 & False/True\\
can\_attack & float & 3 & False/True\\
\hline
\end{tabular}
\begin{tabular}{l l l l}
clock\_feature & type & placeholder & local/global state \\
\hline
time\_step $\quad \quad \quad \quad $ & float & 1 & True/True \\
\hline
\end{tabular}
\end{center}
\end{table}

\textbf{Asynchronous action Space}.
The discrete set of actions that agents are allowed to act consists of movement in six directions (one hexagon per time), shoot[enemy\_id], guide shoot[enemy\_id] and stop.
Agents moving to next hex must cost specific time steps: tank and chariot costing 1 time step and infantry for 5 time steps (show in figure~\ref{yibu}).
This difference cause synchronization.
For chariot and infantry, after arriving destination, if they want to take shoot or guide shoot action, they should keep stop for several specific time steps, and the shoot and guide shoot action only occur when there have been enemies in sight and in shooting range with the shoot cooling time finished.
But for tank, it can shoot while moving, and tank also has shoot cooling time.
Tank can not take guide shoot, but the ability of shoot while moving makes it very powerful.
Instead, even though the infantry and chariot must perform stop before shoot, they can cooperate with each other to complete the guide shoot.

\textbf{Rewards}.
The overall goal is to maximise the win rate for each battle.
The default reward setting is to use a reward dealt by calculating the difference between the loss of allied health and the loss of the enemy's health.

\subsection{Six scenarios}
\begin{figure*}
   \begin{center}
   \includegraphics[width=0.975\textwidth]{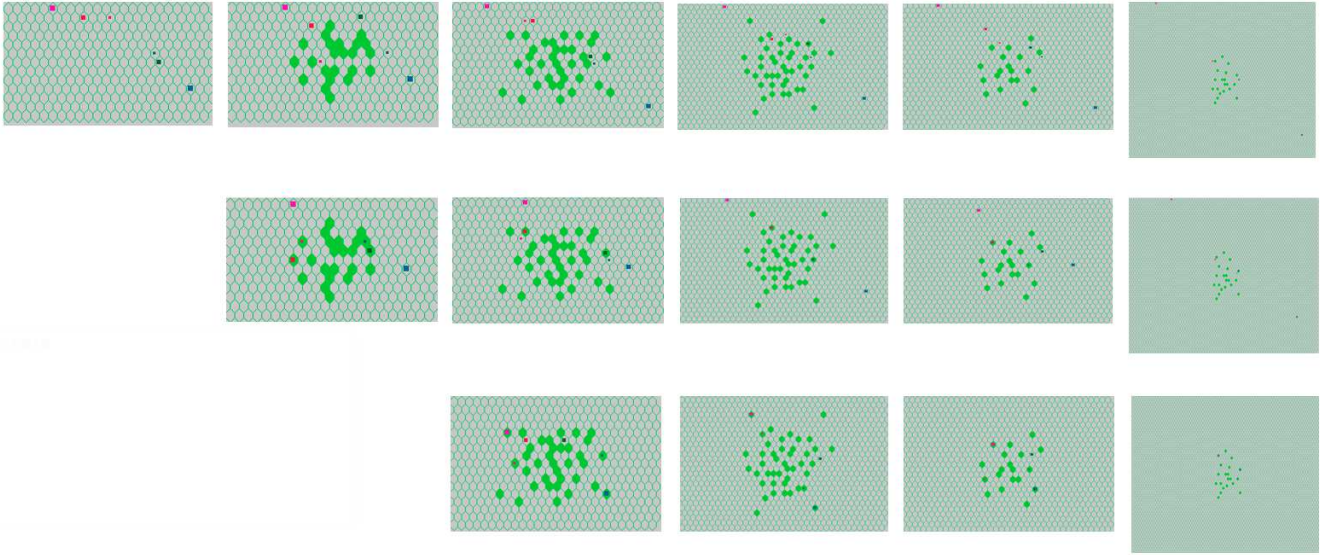}
   \end{center}
   \caption{We show all built-in bots fighting against themselves in all scenarios. The first, second and third lines are KAI0, KAI1 and KAI2 in all scenarios, respectively. We can see KAI0 rush to center of the battlefield, and if there has been special terrain in map, KAI0's operators may go through the special terrain without staying too much. KAI1's operators pay more attention on special terrain, and the tank and the chariot always cooperate together nearby the special terrain. KAI2's chariot and infantry will go near the special terrain and hide for guiding shoot, and tank will rush to the front to attract firepower.}
   \label{logo1}
\end{figure*}

Along with our published environment, we release six scenarios as basic benchmarks for developing asynchronous MARL algorithms.
Operator details of different scenarios are shown in Table \ref{tab:pro}, and screenshots are displayed in Figure \ref{logo1}.

\begin{table*}[htbp]
\begin{center}
\caption{Operators attributes in different scenarios.}
\label{tab:pro}
\begin{tabular}{lllll}
\hline
scenarios & map size & operators  & special terrain & guide shoot \\
\hline
scenario 0 & (13,23) &  tank, chariot and infantry in red and blue teams& False & False \\
scenario 1 & (13,23) &  tank, chariot and infantry in red and blue teams & True & False\\
scenario 2 & (17,27) & tank, chariot and infantry in red and blue teams & True & True\\
scenario 3 & (27,37) & tank, chariot and infantry in red and blue teams & True & True\\
scenario 4 & (27,37) & tank, chariot and infantry in red and blue teams & True & True\\
scenario 5 & (67,77) & tank, chariot and infantry in red and blue teams & True & True\\
\hline
\end{tabular}
\end{center}
\end{table*}

\begin{figure}
   \begin{center}
   \includegraphics[width=0.4\textwidth]{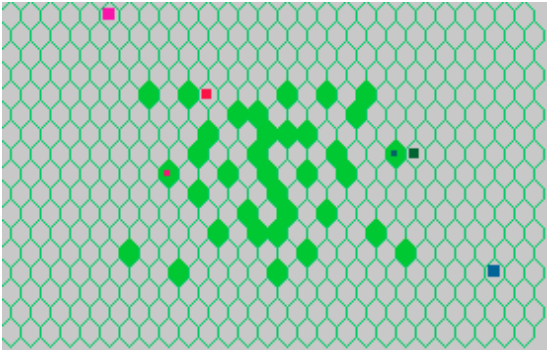}
   \end{center}
   \caption{KAI2 red fight against KAI2 blue in scenario 3.}
   \label{bot}
\end{figure}

\textbf{Built-in Opponent Bots}.
POCA offers built-in Bot mode, where the RL agents can fight with the built-in rule-based bots with different strategies:
\begin{itemize}
\item KAI0: a rush AI with the highest attack priority and directly going to the center of the battle field.
\item KAI1: an AI keeps hiding to ambush near special terrain, which is a commonly used strategy in Wargames.
\item KAI2: an AI uses guided shoot, i.e., let operators squat on special terrain for a long time and attack enemies with guide shoot because of the advantage of sight, and an example is shown in Figure \ref{bot}.
\end{itemize}

\textbf{Control modes}.
Along with the above Built-in bots model, POCA provides another two control modes.
The first one is Self-Play mode for reinforcement learning.
The second mode is the human mode, and human players can use POCA for fighting against the built-in bots, RL agents and other human players.

\begin{figure}
   \begin{center}
   \includegraphics[width=0.45\textwidth]{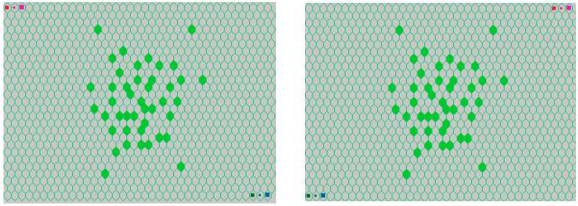}
   \end{center}
   \caption{How changes of Json file impact scenario.}
   \label{gai_hex}
\end{figure}

\begin{figure}
   \begin{center}
   \includegraphics[width=0.45\textwidth]{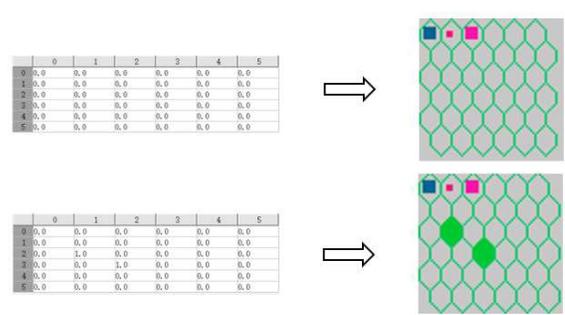}
   \end{center}
   \caption{How changes of hdf5 file impact map.}
   \label{mapedit}
\end{figure}

\textbf{Environment modification}.
POCA also provides function to modify the map and game settings by editing json file in code.
For example, edit "initHex" can change the position of operator at the beginning of the game, as shown in figure~\ref{gai_hex}.
Specifically, by modifying other attribute settings such as move speed, POCA can degenerate to a synchronous cooperation game.
Moreover, a map can be changed by editing the hdf5 file as shown in figure~\ref{mapedit}.
Details can be found in our released code.

\begin{figure}
   \begin{center}
   \includegraphics[width=0.475\textwidth]{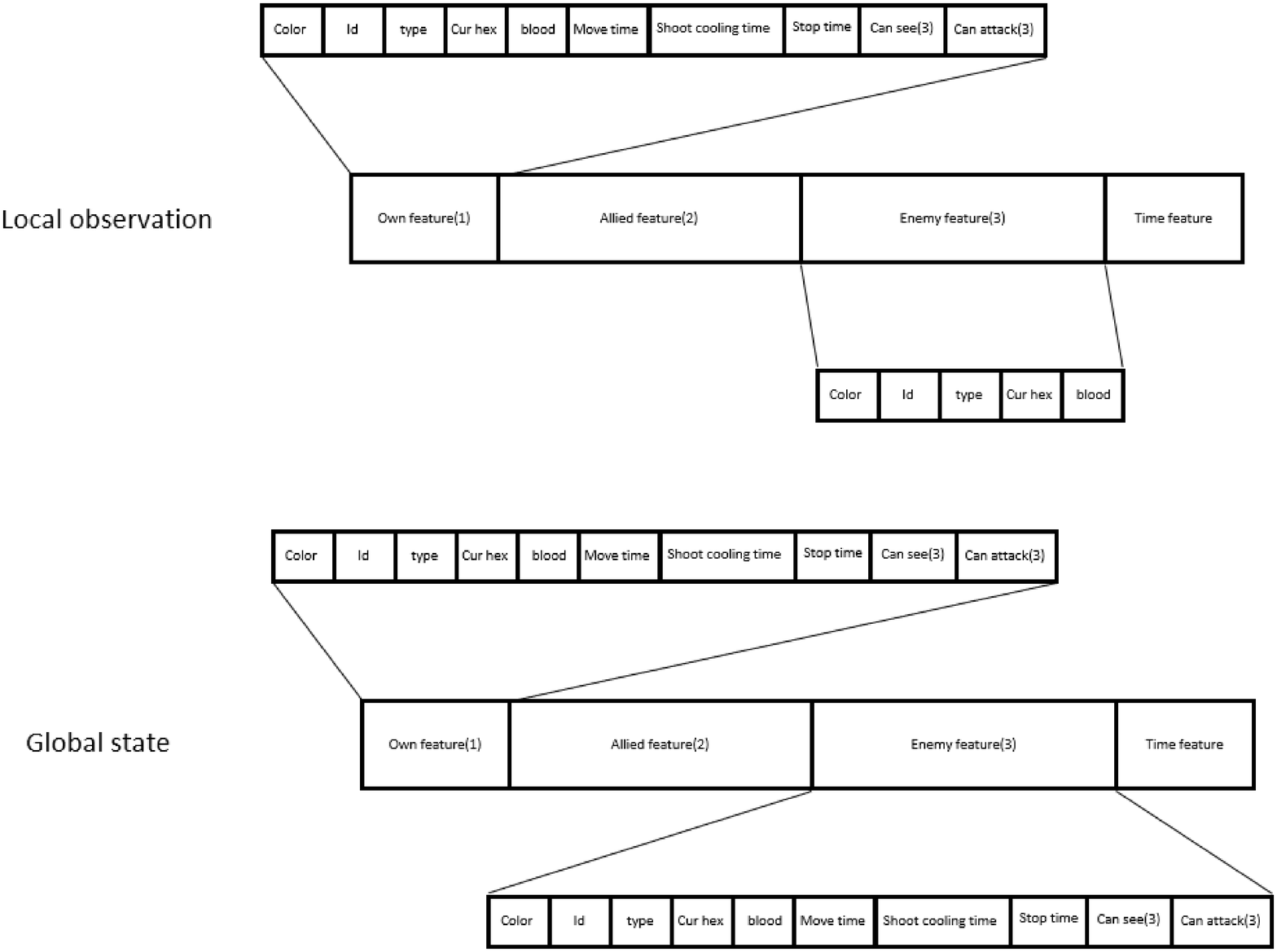}
   \end{center}
   \caption{Feature vectors in all the experiments.}
   \label{fig:featureorg}
\end{figure}

\begin{figure*}
   \begin{center}
   \includegraphics[width=0.95\textwidth]{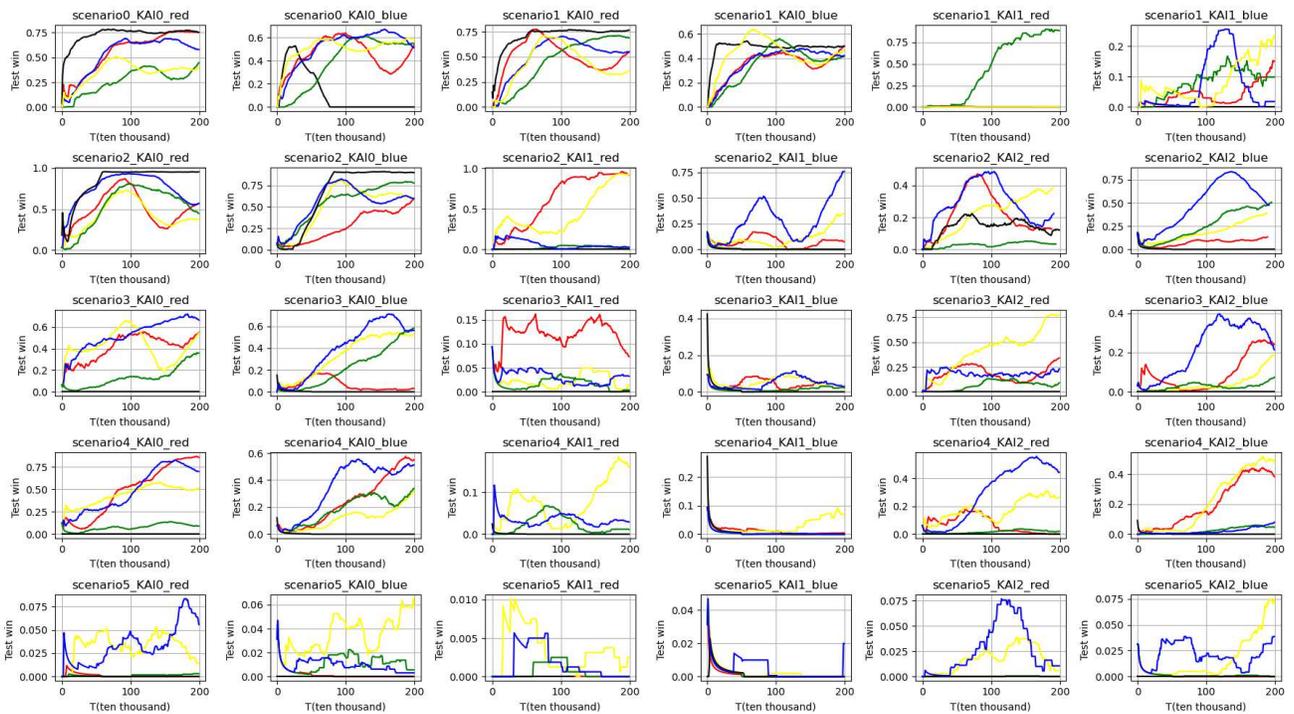}
   \end{center}
   \caption{Results of MARL methods on POCA tasks. Red lines are QMIX, blue lines are IQL, yellow lines are VDN, green lines are qtran, and black lines are coma. scenario0-KAI0-red represents MARL agent as red team training with KAI0 as opponent in scenario 0.}
   \label{logo11}
\end{figure*}

\section{Experiments}
\subsection{Settings}
In this section, we present experiments on POCA tasks of six scenarios.
The purpose of these experiments is to provide baselines for future study, and highlight how changes of scenarios challenge existing popular MARL methods.
All experiments use the default rewards throughout all scenarios, as introduced in previous section.
The agent local observations used in the experiments include features in table~\ref{tab:feature}, and we organize the feature to be a vector as shown in figure~\ref{fig:featureorg}.
As for evaluation, the training is paused after every 10000 time steps during which 32 test episodes are run with agents performing action selection greedily in a decentralised fashion.
The percentage of wining episodes is reported, where the agents' blood is larger than enemies within 600 time steps is referred as win.
The architecture of each agent network is a DRQN with one recurrent layer.
Specifically, the network consits of a GRU with a 64-dimensional hidden state, and a fully-connected layer before and after.
Our baseline for all tasks is based on PyMARL with similar experiments setting\footnote{https://github.com/oxwhirl/pymarl}.

\subsection{Results and analysis}
\textbf{Built-in bots vs. build-in bots}.
We design three different built-in bots, by using distinct types of strategies.
All the built-in bots can be applied in all scenarios and served as red or blue.
Table \ref{result} shows results of built-in bots combats with 32 running times.
In scenario 0, there is no special terrain, and when KAI0-red fights with KAI0-blue, the win rate is close to 0.5.
In all scenarios, KAI1 has relatively good performance, but KAI2 only performs well in several scenarios.
We guess it is the limited design of using guide shoot, which is alleviated by MARL algorithms in the latter experiments.
Moreover, strategies of the three built-in bots are briefly displayed in Figure \ref{logo1}.
\begin{table}[htbp]
\begin{center}
\caption{Results of built-in bots fighting with each other in all scenarios, and the number xxx/xxx means win rate/average time steps when game finished.}
\label{result}
\begin{tabular}{l l}
scenario 0 & KAI0-blue\\
\hline
KAI0-red & 0.438/284.3\\
\hline
\end{tabular}
\begin{tabular}{l l l}
\hline
scenario 1 & KAI0-blue & KAI1-blue \\
\hline
KAI0-red & 0.438/359.7 & 0.063/510.3\\
KAI1-red & 0.625/510.3 & 0.187/588.1\\
\hline
\end{tabular}
\begin{tabular}{l l l l}
\hline
scenario 2 & KAI0-blue &  KAI1-blue  & KAI2-blue \\
KAI0-red & 0.375/438.0 & 0.282/564.8 & 0.031/591.78\\
KAI1-red & 0.531/501.1 & 0.562/600.0 & 0.593/600.0\\
KAI2-red & 0.625/586.2 & 0.344/600.0 & 0.094/600.0\\
\hline
\end{tabular}
\begin{tabular}{l l l l}
\hline
scenario 3 & KAI0-blue &  KAI1-blue  & KAI2-blue \\
KAI0-red & 0.438/433.3 & 0.031/579.2 & 0.000/600.0\\
KAI1-red & 0.594/548.6 & 0.156/600.0 & 0.781/600.0 \\
KAI2-red & 0.750/600.0 & 0.031/600.0 & 0.218/600.0\\
\hline
\end{tabular}
\begin{tabular}{l l l l}
\hline
scenario 4 & KAI0-blue &  KAI1-blue  & KAI2-blue \\
KAI0-red & 0.406/509.9 & 0.282/538.3 & 0.063/600.0 \\
KAI1-red & 0.594/516.3 & 0.375/582.8 & 0.688/600.0\\
KAI2-red & 0.813/534.9 & 0.125/600.0 & 0.250/600.0\\
\hline
\end{tabular}
\begin{tabular}{l l l l}
\hline
scenario 5 & KAI0-blue &  KAI1-blue  & KAI2-blue \\
KAI0-red & 0.344/531.3 & 0.125/570.6 & 0.031/597.9\\
KAI1-red & 0.781/579.2 & 0.063/600.0 & 0.875/600.0\\
KAI2-red & 0.469/585.7 & 0.188/600.0 & 0.375/600.0\\
\hline
\end{tabular}
\end{center}
\end{table}

\textbf{Multi-agent reinforcement learning bots}.
For MARL experiments, i.e., QMIX\cite{rashid2018qmix}, QTRAN\cite{son2019qtran}, COMA\cite{foerster2018counterfactual}, VDN\cite{sunehag2017value} and IQL\cite{tampuu2017multiagent}, we use their original algorithms to perform asynchronous cooperation by using empty action when no executed actions can be obtained for an agent.
we choose results of best seeds, and the results are shown in Figure \ref{logo11}.
We can see most agents can perform well when fighting against KAI0, but in some harder scenarios like in sceniro5, none of the agents can win more than 10\%.
As for opponents KAI1 and KAI2, almost all the agents cannot perform well.
Based on the results, we believe POAC with the built-in bots can serve as a perfect benchmark for MARL algorithms under asynchronous cooperation settings.

\begin{figure}
   \begin{center}
   \includegraphics[width=0.3\textwidth]{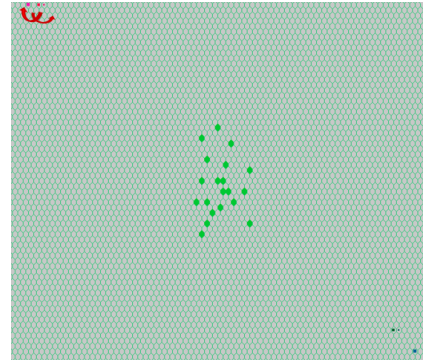}
   \end{center}
   \caption{Agents move back and forth in the position of the arrow, showing no effective strategies learned.}
   \label{fupan}
\end{figure}

\begin{figure}
   \begin{center}
   \includegraphics[width=0.495\textwidth]{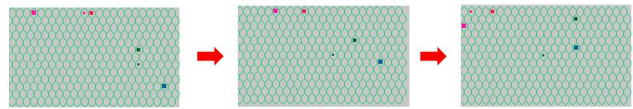}
   \end{center}
   \caption{The red agents will meet the enemy first, then fight against the enemy, and finally retreat to the origin, focusing on attacking the single enemy. Overall, the red agents successfully learn to use vision.}
   \label{fig:fupanwin}
\end{figure}

From the training and testing replays, we find almost all algorithms can learn strategy of guide shoot in smaller sized maps, and this may be the reason for improvement of win rate.
In scenarios 4 and 5, all algorithms perform bad, and we review some replays, as show in Figures \ref{fupan} and \ref{fig:fupanwin}, indicating success and failure cases, respectively.
We believe researchers can analyse and improve their MARL algorithms through our provided replay function.


\section{Conclusion}
In this paper, we have presented POAC as a benchmark for partially observable asynchronous multi-agent cooperation challenge.
Inspired by Wargame, POAC provides six diverse combat scenarios with different types of built-in bots, which largely challenge current MARL methods.
In the near future, we aim to extend POAC with more challenging scenarios such as increasing numbers of operators and rules.
We look forward to accepting contributions from the community to POCA and also hope it will become a standard benchmark for measuring progress in cooperative MARL.

\bibliographystyle{IEEEtran}
\bibliography{ref}

\end{document}